%   TeX file 
%   Super-Eddington Radiation Transfer in Soft Gamma Repeaters
%   by Andrew Ulmer
\magnification=\magstep1
\headline={\ifnum\pageno=1\hfil\else\hfil\tenrm--\ \folio\ --\hfil\fi}
\footline={\hfil}
\hsize=6.0truein
\vsize=8.54truein
\hoffset=0.25truein
\voffset=0.25truein
\baselineskip=20pt
%
% No hyphenation
%
\tolerance=9000
\hyphenpenalty=10000
%
% FONTS
%
% smcap is small capitals for lower case characters.
% Used in authors name, postal addr., and table headings if desired.
%

% a boldface math font
\font\mbf=cmmib10 \font\mbfs=cmmib10 scaled 833
% a boldface math symbol font
\font\msybf=cmbsy10 \font\msybfs=cmbsy10 scaled 833

%
% MATH MACROS
%
% Using exercise 17.20 to define the bold face math symbols
% Others can be added as needed.
% Use in this way ${\bmsy\grad\cdot}{\bmit\xi} =0$. "div \xi=0"
%

\textfont9=\mbf \scriptfont9=\mbfs \scriptscriptfont9=\mbfs

\textfont10=\msybf \scriptfont10=\msybfs \scriptscriptfont10=\msybfs
%
% Define Greek letters to allow use when \mbf is active font
%
\mathchardef\alpha="710B
\mathchardef\beta="710C
\mathchardef\gamma="710D
\mathchardef\delta="710E
\mathchardef\epsilon="710F
\mathchardef\zeta="7110
\mathchardef\eta="7111
\mathchardef\theta="7112
\mathchardef\iota="7113
\mathchardef\kappa="7114
\mathchardef\lambda="7115
\mathchardef\mu="7116
\mathchardef\nu="7117
\mathchardef\xi="7118
\mathchardef\pi="7119
\mathchardef\rho="711A
\mathchardef\sigma="711B
\mathchardef\tau="711C
\mathchardef\upsilon="711D
\mathchardef\phi="711E
\mathchardef\chi="711F
\mathchardef\psi="7120
\mathchardef\omega="7121
\mathchardef\varepsilon="7122
\mathchardef\vartheta="7123
\mathchardef\varpi="7124
\mathchardef\varrho="7125
\mathchardef\varsigma="7126
\mathchardef\varphi="7127
%define \nabla to allow use when \msybf is active font
\mathchardef\nabla="7272
%define \cdot to allow use when \msybf is active font
\mathchardef\cdot="7201
 % rename \nabla as \grad
%
%
% \ldb and \rdb are double brackets, for use in formulae
%

%
% \bigldb and \bigrdb are \big size double brackets
%

%
% \biggldb and \biggrdb are \bigg size double brackets
%

%
% \arcsec produces arcsec symbol so that 3\arcsec5 produces 3."5 with the
% second symbol and the period aligned.
%

%
% EQUATION NUMBERING MACROS
%
\newcount\eqnumber
\eqnumber=1
% \new macro produces sequentially numbered equations by writing \eqno(\new)
% at end of displayed equations
%
\def\new{{\the\eqnumber}\global\advance\eqnumber by 1}
%
% To refer to an equation which is 5 equations back, write "equation (\ref5)"
%
\def\ref#1{\advance\eqnumber by -#1 \the\eqnumber
     \advance\eqnumber by #1 }
%
% \last macro is like \new except counter is not advanced. Useful for
% equations which are in parts a and b.
%
\def\last{\advance\eqnumber by -1 {\the\eqnumber}\advance 
     \eqnumber by 1}
%
% To name an equation, place command "\eqnam{\Poisson}" before equation, and
% thereafter "equation(\Poisson)" will generate the proper equation number.
%
\def\eqnam#1{\xdef#1{\the\eqnumber}}
%
% Reference macros
%
% To generate reference to a paper in Ap.J. volume 300, p.123 
% write \apj{Claus, S. 1990}{300}{123}
%

%
% Define macros to start sections or subsections of the paper
% e.g.  \sect{1.~INTRODUCTION}
% e.g.  \subsec{2.2}{Mode Classification}
\def\sect#1 {
  \bigbreak
  \centerline{\bf #1} 
 \bigskip}
\def\subsec#1#2 {
  \bigbreak
  \centerline{#1.~{\it #2}}
  \bigskip}
% Definition of \figure: Includes advancing the counter (figno) at each call.
\newcount\figno
\figno=0
\def\figure{\global\advance\figno by 1 Figure~\the\figno.~}
%======================================================
%
% Some useful symbols
%

\def\>{$>$}
\def\<{$<$}

\def\simlt{\lower.5ex\hbox{$\; \buildrel < \over \sim \;$}}
\def\simgt{\lower.5ex\hbox{$\; \buildrel > \over \sim \;$}}
\def\sqr#1#2{{\vcenter{\hrule height.#2pt
      \hbox{\vrule width.#2pt height#1pt \kern#1pt
         \vrule width.#2pt}
      \hrule height.#2pt}}}

%
%  macro for invoking today's date (when TeX is run on your file)
%

\def\today{\ifcase\month\or
	January\or February\or March\or April\or May\or June\or
	July\or August\or Setrueptember\or October\or November\or December\fi
	\space\number\day, \number\year}
%  define heading for your pages.
%
\def\head#1{\headline={\ifnum\pageno>1
	{\tenrm #1} \hfil Page \folio
	\else\hfil\fi}}
%
%  This provides a simple means of producing references in ApJ style.
\def\ref #1;#2;#3;#4{\par\pp #1, {\it #2}, {\bf #3}, #4}
\def\book #1;#2;#3{\par\pp #1, {\it #2}, #3}
\def\rep #1;#2;#3{\par\pp #1, #2, #3}
%   \ref will construct a reference in Ap.J style.  The first 3 arguments
%           must be delimited by semicolons,  and the last by a space.
%           all four arguments must be supplied, e.g.
%        
%   \ref  Blow, Joe, 1917;J.I.R.;1;991
% 

% \gax is a math symbol with a '>' over '\sim' (a sqiggle), and
% \lax is a math symbol with a '<' over '\sim' (a sqiggle),
\newbox\grsign \setbox\grsign=\hbox{$>$}
\newdimen\grdimen \grdimen=\ht\grsign
\newbox\laxbox \newbox\gaxbox
\setbox\gaxbox=\hbox{\raise.5ex\hbox{$>$}\llap
     {\lower.5ex\hbox{$\sim$}}}\ht1=\grdimen\dp1=0pt
\setbox\laxbox=\hbox{\raise.5ex\hbox{$<$}\llap
     {\lower.5ex\hbox{$\sim$}}}\ht2=\grdimen\dp2=0pt

\def\etal{{\it et~al.\ }}
%%%%  end of input of apjmacros.tex

\def\paczy{Paczy{\'n}ski}

\def\Epar{E_{\|}}
\def\Eper{E_{\perp}}
\def\pd{\partial} 
%
%  \AUtoday

%
\centerline{\bf Super-Eddington Radiation Transfer in Soft Gamma Repeaters}
\vskip 0.5cm
\vskip 0.5cm
\centerline{Andrew Ulmer}
\centerline{Princeton University Observatory, Princeton NJ 08544}
\vskip 0.5cm

%\centerline{Submitted to Astrophysical Journal Letters}
%\vskip 0.5cm

\centerline{\bf ABSTRACT}
Bursts from soft gamma repeaters have been shown to be super-Eddington by a
factor of 1000 and have been persuasively associated with compact objects.
Here, a model of super-Eddington radiation transfer on the surface of a
strongly magnetic ($\geq 10^{13}$ gauss) neutron star is studied and related
to the observational constraints on soft gamma repeaters. In strong magnetic
fields, the cross-section to electron scattering is strongly suppressed in one
polarization state, so  super-Eddington fluxes can be radiated while the
plasma remains in hydrostatic equilibrium. The model offers a somewhat natural
explanation for the observation of similarity between spectra from bursts of
varying intensity. The radiation produced in the model is found to be linearly
polarized to about 1 part in 1000 in a direction determined by the local
magnetic field, and the large intensity variations between bursts are
understood as a change in the radiating area on the source. Therefore, the
polarization may vary as a function of burst intensity, since the complex
structure of the magnetic field may be more apparent for larger radiating
areas. It is shown that for radiation transfer calculations in this limit of
super-strong magnetic fields it is sufficient to solve the radiation transfer
equations for the low opacity state rather than the coupled equations for both.
With this approximation, standard stellar atmosphere techniques are utilized
to calculate the model energy spectrum.

\medskip
{\it Subject Headings:} Gamma-Rays: Bursts -- Radiation Processes: Thermal --
Stars: Neutron -- X-Rays: Bursts
\vfill
\eject

\sect{I. INTRODUCTION}

Soft gamma repeaters are a class of super-Eddington, repeating,
high energy transients.
The first repeating soft gamma ray
bursters were recognized in the early eighties: SGR0526-22, (Mazets
and Golenetskii 1981; Golenetskii, Iiyinskii, and Mazets 1984) and
SGR1900+14 (Mazets, Golenetskii, and Guryan 1979).
The discovery of over 100 repetitions of the third repeater,
SGR1806-20 (Laros \etal 1986, 1987; Atteia \etal 1987; cataloged in Ulmer \etal
1993), secured the
claim that soft gamma repeaters are a rare class of transients
distinct from other high energy transients such as x-ray or
gamma-ray bursts. Kouvelioutou \etal (1993, 1994)
have recently observed additional repetitions from the sources SGR1900+14 and
SGR1806-20 with the Burst and Transient Source Experiment.
Properties of SGRs are reviewed in Norris \etal (1991).
Briefly, soft gamma-ray bursts typically last hundreds of milliseconds and have
sharp rise and decay times (Atteia \etal 1987). They have
no simple, discernible pattern of recurrence although they are
clustered in time (Laros \etal 1987). A typical photon energy is
20--30 keV, and there is a strong rollover
below about 15 keV (Fenimore, Laros, and Ulmer 1994).
The sharp rise times as well as the eight second periodicity detected in
the March fifth event (e.g. Mazets \etal 1979) from SGR0526-22 suggest
that the sources are compact objects.

There has been much speculation
as to the distance of the repeaters (e.g. Kouvelioutou \etal 1987;
Norris \etal 1991), particularly because SGR0526-22
is located in the direction of the Supernova remnant N49 in the Large
Magellanic Cloud (Evans \etal 1980) with a very small error box of
0.1 arcmin$^2$ (Cline \etal 1982).
Recent evidence strongly favors a Population~I
distribution. Murakami \etal (1994) corroborated the association
between SGR1806-20 and a plerionic supernova remnant (Kulkarni and
Frail 1993, hereafter KF) by fortuitously imaging a
soft gamma ray burst with the ASCA x-ray satellite and thereby
localizing the burster to within $\sim 8$ arcmin$^2$ --- an
improvement by  a factor of 50 from the previous gamma-ray localization.

The distance to the supernova remnant has been estimated
to be 17 kpc (KF), but, as KF point out,
the distance method (surface-brightness/diameter),
can have large errors (factors of 2 or 3)
due to fluctuations in the local density and magnetic field (e.g. Miln 1979).
In any case, it is likely that the source is at least 5 kpc away.
Fenimore, Laros, and Ulmer (1994) (hereafter FLU), find
that the x-ray spectrum of the bursts from SGR1806-20
has a sharp rollover below
15 keV and can therefore estimate the total flux of an individual burst
by way of analysis of the spectra of 95 soft gamma-ray bursts detected by
the {\it International Cometary Explorer} (ICE).
A lower bound on the flux for the brightest events, assuming a distance of
5 kpc, is $\sim 1.6 \times 10^{41}$~erg~sec$^{-1}$,
$\sim 2 \times 10^{3}$ times the Eddington limit for a neutron star, and
if at a distance of 17 kpc the flux would be larger by a factor of 10.
The repeating bursts from SGR0526-22 were also super-Eddington by a
large factor of $1-2 \times 10^{4}$
assuming a location in the Large Magellanic Cloud.
Furthermore, while the intensity of the individual bursts from
SGR1806-20 has been detected to
vary by over a factor of 50 (Laros, \etal 1987),
the shape of the energy spectrum is remarkably constant
above $\sim 30$~keV and appears to be so at lower energies, too (FLU).

Therefore, the main points that need to be addressed in any
radiation mechanism are that in the context of compact objects,
the mechanism operate at highly super-Eddington fluxes,
that it produce a low-energy roll-over,
and that it produce a similar spectral shape over a wide range of intensities.
Many radiation mechanisms which are partially successful with regard to these
constraints are discussed in FLU.
In \S II, the known mechanisms for producing super-Eddington fluxes
are briefly discussed.
A model which addresses the afore mentioned points
is presented in \S III.
Energy spectra and radiation pressure produced by the model are
calculated in \S IV.
Lastly, \S V contains a discussion of the implications with regard
to observations.

\sect{ II. Super-Eddington Fluxes}

The problem of radiating super-Eddington fluxes from a compact object
has been addressed in only a few different ways.
The Eddington flux level is given by
$$
L_E \sim {4\pi G M m_H c \over \sigma_{Th}}
{\rm ~~erg ~~sec^{-1}}  \eqno(\new)
$$
where $G$ is the gravitational constant, $M$ is the mass of the star,
$m_H$ is the mass of a proton,
$c$ is the speed of light,
and $\sigma_{Th}$ is the Thompson cross section.
In general, there are three paths to tread with regard to the super-Eddington
flux problem:
(1) in some circumstances, the physical cross-section, $\sigma$, is reduced
and the Eddington limit increases so
seemingly super-Eddington fluxes can
be radiated while maintaining hydrostatic equilibrium
(2) the restraining force on the matter is
increased above that of gravity alone, for instance, by
magnetic pressure,
so that the Eddington limit is increased, or
(3) the matter is blown away at relativistic speeds, and a
super-Eddington ``fireball'' is formed in front of it.

The first scenario, which is elaborated in \S III,
has been considered by \paczy ~(1992), who observed that
super-Eddington fluxes may be achievable
in super strong magnetic fields (e.g. Thompson \& Duncan 1993)
where the Thompson cross-section is
suppressed in one polarization state (Herold 1979).

The second scenario has primarily been considered
with respect to gamma-ray burst models (e.g. Lamb 1982) and
generally relies on magnetic pressure to confine the radiating plasma.
For blackbody emission, the required field is
$$
B_{12} > \left[T\over 170 ~{\rm keV} \right]^2.
$$
While the typical temperatures of soft gamma-ray bursts are low enough
that magnetic pressure can counteract the radiation pressure, the
magnetic pressure acts perpendicular to the magnetic field, so matter
is free to move along field lines and will at great speeds unless the
magnetic field is exactly perpendicular to the radiation.
Such a scenario then requires
closed field line geometries.
Also, for dipole fields, the combination
of the radiation and magnetic pressures will tend to slide the radiating
matter towards the equator and away from the star where the magnetic
field is weaker. Such effects allow a smaller region to radiate and
concentrate the matter at the weakest point in the magnetic fields.
These general problems are significant and have yet to be addressed.

Finally, though never seriously considered in the context of
soft-gamma repeaters, many fireball models of gamma-ray bursts are able to
produce super-Eddington fluxes (e.g. \paczy ~1986, Goodman 1986).
The main difficulties in
adapting the fireball ideas to a scenario involving the soft gamma-ray
repeaters stems from the fact that the photon energies are strongly
peaked, quite low, and relatively uniform.
As shown by FLU, over a large range of burst
intensities, the hardness ratio is relatively constant. This means that
for a radiation scenario similar to a blackbody, such as a thermal fireball
model, the area of the fireball must change while maintaining a near constant
temperature. Furthermore, parameters need to be somewhat
finely tuned in order that the energy emerge from a fireball in radiative
rather than kinetic energy (Rees \& M{\'e}z{\'a}ros 1992).
The required fine tuning as well as the problem of changing the area of
a fireball without changing the temperature make such scenarios appear
unlikely, though they have not yet been investigated in detail.

\sect{III. Model of Radiation Transfer}

As discussed above, a radiation mechanism for soft-gamma repeaters must
address the super-Eddington flux problem, provide a strong roll-over in the
data, and be able to produce a range of intensities without appreciatively
changing the energy spectrum. If the sources for soft gamma-ray bursts are
neutron stars with strong magnetic fields of order $10^{13}-10^{14}$~~gauss,
then all of these conditions can be met. In particular, hydrostatic
equilibrium can be maintained as a result of a magnetically suppressed
cross-section. A sharp rollover is produced which results from both
the self absorption of the radiating plasma and the frequency dependent
opacities. If the size of radiating surface varies, a range of
intensities can be generated. Such variations might be produced by the
deposit of a characteristic energy density below the surface which may
occur during glitches or starquakes
(e.g. Epstein 1992 and references therein).
The surface temperature produced by such an energy release would be
a slowly changing function of the depth of release, so that
only a small range of surface temperatures would be observed, in accord with
the observations.
The energy is assumed to be released at large optical depth.
For a neutron star, this condition is met 
if the energy release occurs more than about
a centimeter below the neutron star surface for the low opacity state.

In strong magnetic fields, photon propagation becomes a strong
function of polarization state (e.g. Herold 1979).
In particular, since the motion of electrons is restricted perpendicular to the
magnetic field, the Thompson cross-section for photons with linear polarization
$\Eper$ with
respect to the magnetic field direction (that is, with the plane of the
photon electric field perpendicular to the stellar magnetic field direction)
is much reduced relative to the
normal cross-section.
Herold (1979) gives the relations for the total cross
sections as:
$$
\sigma_{\rm Tot}(\|) \approx \sigma_{Th}\left[ \sin^2\theta + \left(
{\omega \over
\omega_{\rm B}}\right)^2\cos^2\theta \right]  \eqno(\new)
$$
$$
\sigma_{\rm Tot}(\perp) \approx \sigma_{Th}\left( {\omega \over
\omega_{\rm B}} \right)^2  \eqno(\new)
$$
where $\omega$ is the photon frequency,
$\omega_{\rm B} \approx 10 \rm{B}_{12} ~keV$
is the electron cyclotron frequency in the magnetic field and $\theta$ is the
angle between the magnetic field and the Poynting vector.
This equation is a good approximation when $\omega$ is less than
$\omega_{\rm B}$ and larger than $\omega_{\rm B}$/1836, the proton cyclotron
frequency.
In superstrong magnetic fields,
$$
{ \omega_{\rm B}  \over \omega_{\rm Teff}} \approx 20-200
\eqno(\new)
$$
where, $\omega_{\rm Teff}$ is the frequency corresponding to the effective
temperature of soft gamma-ray bursts (5-10 keV generally), so that the
$\Eper$ state has a lower opacity by $400-4\times 10^4$.
In this regime, there is also a simple relation between the differential
cross-sections:
$$
{\sigma(\perp \rightarrow \perp) \over \sigma(\perp \rightarrow \|)} = 3 
  \eqno(\new)
$$
$$
{\sigma(\| \rightarrow \|) \over \sigma(\| \rightarrow \perp)} \sim 
\left( {\omega_{\rm B} \over \omega} \right)^2.   \eqno(\new)
$$

Figure 1 shows a schematic of the radiation transfer under these
circumstances. Due to the lower cross section in the $\Eper$ state,
the mean free path is much longer and energy is transferred to the
surface primarily in this state. Occasionally, a photon will scatter
between states as shown by Eqs. 5,6.
Because of the short timescales
on neutron star surfaces (e.g. light crossing time of 0.1 milliseconds and
Alfv{\'e}n speed order c) and the comparatively long durations of
soft gamma-ray bursts ($\sim 0.5$~~seconds), the matter and two radiation
states have ample time to achieve local thermal equilibrium.
An additional requirement for LTE is a photon generating
radiation process (so that a Wein peak is not formed) which is not
met by the electron scattering processes alone; however
there are a number of second order processes such as double Compton
scattering, proton cyclotron emission, one-dimensional thermal
bremsstrahlung, and photon splitting, which can create photons. 

Therefore, the system is in LTE, and the energy in the system is
transferred primarily in the $\Eper$ state.
The energy transfer can be quantified by examining the flux in the diffusion
limit (e.g. Mihalas 1978):
$$
{F_{\nu} \over 4\pi}
\rightarrow {1 \over 3} {\pd B_\nu \over \pd \tau_\nu}  =
-{1\over \sigma_\nu }{1 \over 3} {\pd B_\nu \over \pd T}{dT \over dz},
  \eqno(\new)
$$
where $\sigma_\nu$ is the cross-section as a function both of frequency and
polarization state. In local thermal equilibrium, the ratio of the fluxes
is given by
$$
{F(\Epar) \over F(\Eper)} \approx {\sigma_{\rm Tot}(\perp) \over 
\sigma_{\rm Tot}(\|)} \approx \left( {\omega_{\rm B} \over \omega}\right)^2
\approx 10^3-10^4.
  \eqno(\new)
$$
The radiation pressure in a static medium is 
$$
P= {1 \over c}\int \sigma_\nu {F} d \nu,
  \eqno(\new)
$$
therefore, the pressure from the $\Eper$ state photons is equal to that
from the $\Epar$ state. The sum of the pressures is less than what would
be found in a non-magnetized system by roughly a factor of
$0.5\times (\omega_{\rm B} / \omega)^2$, so the effective
Eddington limit is much higher (\paczy ~1992).

To calculate the emergent energy spectrum from such a highly magnetized
source, it suffices to follow the flow of radiation in the low opacity
state. In particular, the cross section in the $\Eper$ state
can be separated into a scattering component,
$\sigma_{\rm s}$,
where the photon scatters into the $\Eper$ state
and an absorption component, where the photon scatters into the $\Epar$ state.
The differential cross sections in the $\Epar$ states (Eqs. 5,6)
are such that a photon
will scatter numerous times within the high opacity state, thermalize due
to the photon production processes described above
and eventually scatter back into into the low opacity state.
Double Compton scattering likely plays the dominate role in the thermalization
(photon-production) because it is expected to be roughly a factor of
($\alpha \approx 1/137$) smaller than the magnetically suppressed Thompson
cross-section of Eq. 2 and a photon will likely scatter a thousand or more
times before scattering out of the high opacity state.

\sect{Model Calculations and Results}

With the model described above, the energy spectrum
can be calculated with regard only to the low opacity polarization
state, so that the problem reduces to the calculation of a 
unpolarized stellar atmosphere with cross-sections for
scattering and absorption determined by the suppressed Thompson
scattering relations.
Because the cross-section
is a smooth function of frequency and the scattering and absorption portions
of the cross-section are comparable, many
methods are available for the calculation of the
emergent spectra. Here, an iterative solution to the one-dimensional
radiation transfer
equation is utilized (e.g. Mihalas 1978 (6-1)) and the requirement
of conservation of flux is met to within 2 percent with
iteratively correcting the temperature as a function of optical depth
using a procedure developed by Lucy (Lucy 1964; Mihalas 1978 (7-2)).
The use of angle averaged quantities the iteration procedure
in general is a good approximation, however, the magnetic field direction
introduces a small anisotropy in the cross-sections,
for which a more detailed calculation should account. Note however that
the scattering/absorption ratio  (Eq. 5) is angle independent.

The spectrum resulting from solving the radiation transfer equations
is shown in figure 2.
The emergent flux is strongly altered from a blackbody. The peak flux
occurs at twice the effective temperature (2/3 that of a blackbody).
The spectrum is primarily a function of
$T_{\rm eff}$. There is a much weaker dependence on
$\omega_{\rm B} / \omega$ (100 in figure 2)
when the ratio is in the range of 20-200
since this primarily scales the cross-section without changing its shape.
Therefore, the spectral shape is a good approximation over a range of
$\omega_b/\omega$.

Using the spectrum obtained from the radiation transfer model,
the radiation pressure can be calculated and
characterized as a function of peak flux energy:
$$
P(\Eper)= {1 \over c}\int \sigma_\nu {F_{\nu}(\Epar)} d \nu \approx {2.9\over
c} \sigma_{\rm Th} \sigma_{\rm B} T_{\rm eff}^4 {\omega_{\rm Teff} \over
\omega_{\rm B}}
  \eqno(\new)
$$
where $\sigma_{\rm B}$ is the Stefan-Boltzmann constant.
In the diffusion limit,
where the radiation pressure will generally be greatest, the high opacity
state contributes approximately the same radiation
pressure as the low opacity. Therefore, the Eddington
limit in this regime, expressed as a function of the non-magnetic
Eddington limit is:
$$
L_{\rm Edd} \approx {1\over 6.8} \left[ {\omega_{\rm Teff} \over
\omega_{\rm B}}\right]^2 \approx 60-6000 ~L_{\rm Edd}.
  \eqno(\new)
$$

\sect{Discussion}

Using the spectral shape derived above, the radiating area of the source
can be estimated.
For the events from SGR1806-20 for which complete x-ray spectra
were determined (FLU), the peak flux occurs between about 8 and 20 keV.
The effective temperature is found to be half of the frequency of
peak flux, so
the projected active source area for the brightest burst from SGR1806-20 is
$$
A = {D^2 L_{\rm measured} \over \sigma_{\rm B} T_{\rm eff}^4} \approx
2 \times 10^3 {\rm km}^2 \left(D \over 10 ~{\rm kpc}\right)
\left(T_{\rm eff} \over 7 ~{\rm keV}\right)^{-4},
\eqno(\new)
$$
where the uncertainties in $T_{\rm eff}$ and $D$ can sway the area
by about a factor of 20 in either direction. The maximum projected
radiating area for a neutron star is approximately $\pi r^2 \approx 300$~km,
which is well within the error bars on the determined area. However,
if the distance is much greater than 10 kpc and the
effective temperature is less than about 10 kev, it may be difficult to
reconcile the model with observations.
Blackbody models would generally require even larger
areas ($\sim 5-10$) because for a blackbody, $T_{\rm eff} \approx
T_{\rm peak}/3$.
One obtains similar results for the bursts from SGR0526-22.
The distance to this source is known with some certainty to be about 55 kpc;
however, there were no x-ray observations below 30 keV,
so it is not possible to determine
where the frequency of peak flux occurs for this burster.

The radiation transfer model offers a somewhat natural explanation
for the similarity between spectra of bursts with different
intensities.
If a characteristic energy density, or temperature, is deposited
at large optical depth, $\tau$, within the star, the resulting
surface temperature would only be a weak function of $\tau$. 
At large optical depth, the temperature relation for the
grey atmosphere limit becomes a reasonable approximation, so
$$
T^4 = {3 \over 4}T^4_{\rm eff}[\tau + q(\tau)],
\eqno(\new)
$$
where $q(\tau)$ is a number of order unity.
Inverting the equation shows that for releases of constant energy density,
so that $T$ is constant, the effective surface temperature goes as the
fourth root of the optical depth of release, $\tau$.
However, as Thompson and Duncan (1992) observed,
the suppressed opacity rises with temperature and therefore with optical depth.
If the energy is released quite deep in the crust, the radiation pressure
may be higher than the Eddington limit if, for example, the typical temperature
is near the electron cyclotron energy.
However, at such large optical depth, pressure from the matter above could
offer a sufficient restraining force to prevent expansion.

Beyond meeting the current observational constraints,
this radiation transfer model predicts that SGR bursts should be
locally linearly polarized to approximately 1
part in 1000, because the flux is dominated by the low opacity state.
Polarization may then be a function of burst intensity,
since the destruction of the polarization pattern by the global structure
of the magnetic field become more apparent for larger intensities and areas.
Second, if the variations in intensity of the soft gamma-ray bursts
are, indeed, a result of varying sizes the ``active'' region, then
there should be a critical intensity reached when the
entire source is active. 

{\it Acknowledgments}
I wish to thank R.~I.~Epstein, E.~E.~Fenimore, and B.~Paczy{\'n}ski for
helpful discussions.
This work was supported by NASA Grant NAG5-1901.

\vfill\eject

\sect{REFERENCES}

\par\noindent\hangindent=3pc\hangafter=1
Atteia, J. L., \etal 1987 Ap. J. (Letters) { 320}, L105

\par\noindent\hangindent=3pc\hangafter=1
Fenimore, E.E., Laros, J.G., Ulmer A., (FLU),  ApJ, in press.

\par\noindent\hangindent=3pc\hangafter=1
Goodman, J. 1986, Ap.J. (Letters) 308, L47

\par\noindent\hangindent=3pc\hangafter=1
Epstein, R. I. 1992 in
{\it Gamma-Ray Bursts Observations, Analyses
and Theories} eds C. Ho, R. I. Epstein, and E. E. Fenimore (Cambridge:
Cambridge University Press) p 1

\par\noindent\hangindent=3pc\hangafter=1
Golenetskii, S.V., Aptekar, R.L., Guran, Yu.A., Iiyinskii, V.N., and 
Mazets, E. P. 1987, Soviet Astron. Lett., 13, 166

\par\noindent\hangindent=3pc\hangafter=1
Kouveliotou, C., \etal 1987 Ap. J. (Letters) { 322}, L21

\par\noindent\hangindent=3pc\hangafter=1
Kouveliotou, C., \etal 1993 Nature { 362}, 728

\par\noindent\hangindent=3pc\hangafter=1
Kouveliotou, C., \etal 1994 Nature { 368}, 125-127.

\par\noindent\hangindent=3pc\hangafter=1
Kulkarni, S. R., and Frail, D. A., (KF) 1993 Nature { 365} 33-35.

\par\noindent\hangindent=3pc\hangafter=1
Laros, J. G., \etal 1986, Nature, { 322}, 152-153.

\par\noindent\hangindent=3pc\hangafter=1
Laros, J. G., \etal 1987, Ap. J. (Letters), { 320}, L111

\par\noindent\hangindent=3pc\hangafter=1
Lucy, L., 1964 in
{\it Harvard Smithsonian Conference on Stellar Atmospheres:
Proceedings of the First Conference}, S.A.O. report 167, (Cambridge:
Smithsonian Astrophysical Society) p 93

\par\noindent\hangindent=3pc\hangafter=1
Mazets, E. P., Golenetskii,
 S. V., Iiyinskii,
 V. N., Aptekar, R. L., and Guryan, Y. A., 1979, Nature, { 282} 587

\par\noindent\hangindent=3pc\hangafter=1
Mazets, E. P., Goleneskii, S. V., Guryan, Y. A., 1979
Sov. Astron Let, 5(6), Nov-Dec,  343

\par\noindent\hangindent=3pc\hangafter=1
Mazets, E. P. and Golenetskii, S. V., 1981 Astrophys. Space Sci. { 75} 47

\par\noindent\hangindent=3pc\hangafter=1
Mihalas, D., 1978 {\it Stellar Atmospheres}, (San Francisco:
W.H. Freeman and Co.)

\par\noindent\hangindent=3pc\hangafter=1
Miln, D.K., 1979 Aust. J. Phys. 32, 83

\par\noindent\hangindent=3pc\hangafter=1
Murakami, T., \etal , 1994 Nature, 368, 127

\par\noindent\hangindent=3pc\hangafter=1
Paczy{\'n}ski, B. 1992, Acta Astronomica { 42} 145

\par\noindent\hangindent=3pc\hangafter=1
Paczy{\'n}ski, B. 1986, Ap. J. (Letters) 308, L43

\par\noindent\hangindent=3pc\hangafter=1
Rees \& M{\'e}z{\'a}ros 1992, MNRAS 258, 41P

\par\noindent\hangindent=3pc\hangafter=1
Thompson, C., Duncan, R.C. 1993, Ap.J. 408, 194

\par\noindent\hangindent=3pc\hangafter=1
Ulmer, A., Fenimore, E. E., Epstein, R. I., Ho, C., Klebesadel, R. W.,
Laros, J. G., and Delgato, F. 1993, Ap. J. 418, 395

\vfill\eject

\sect{FIGURE CAPTIONS}

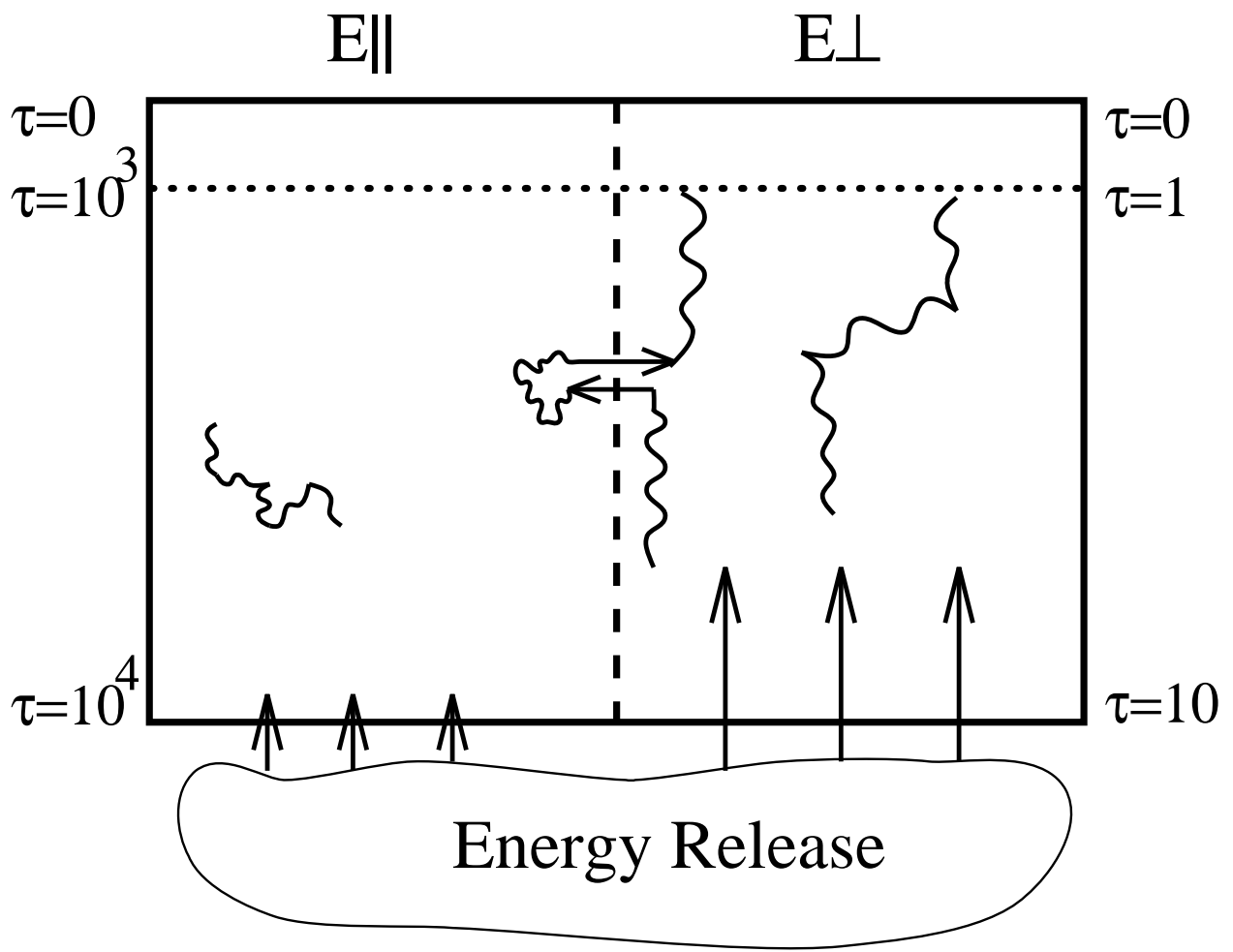
{\bf Fig.~1}: This schematic illustrates the major properties of
the radiation transfer in a strongly magnetized medium.
The electron scattering cross-sections for photons differ by
about a factor of 1000 between the two photon polarization
states $\Eper$ and $\Epar$ 
(plane of the photon electric field perpendicular or parallel to
the magnetic field direction). Consequently, the flux in the low
opacity,
$\Eper$ state is much higher, and one can see much deeper into the
star in this state.

\vskip 0.5 in
{\bf Fig.~2}: The energy spectra produced in a strongly magnetized
stellar atmosphere is compared to a blackbody. The spectra produced
by the strongly magnetized plasma peaks at a lower frequency and
has a slightly faster fall off at high energies. The shift to lower
energies is produced because the cross-section goes as the square
of the frequency, so that at low frequencies the cross-section is
lower, and one can see farther into the source where the temperatures
are higher.

\vfill\eject
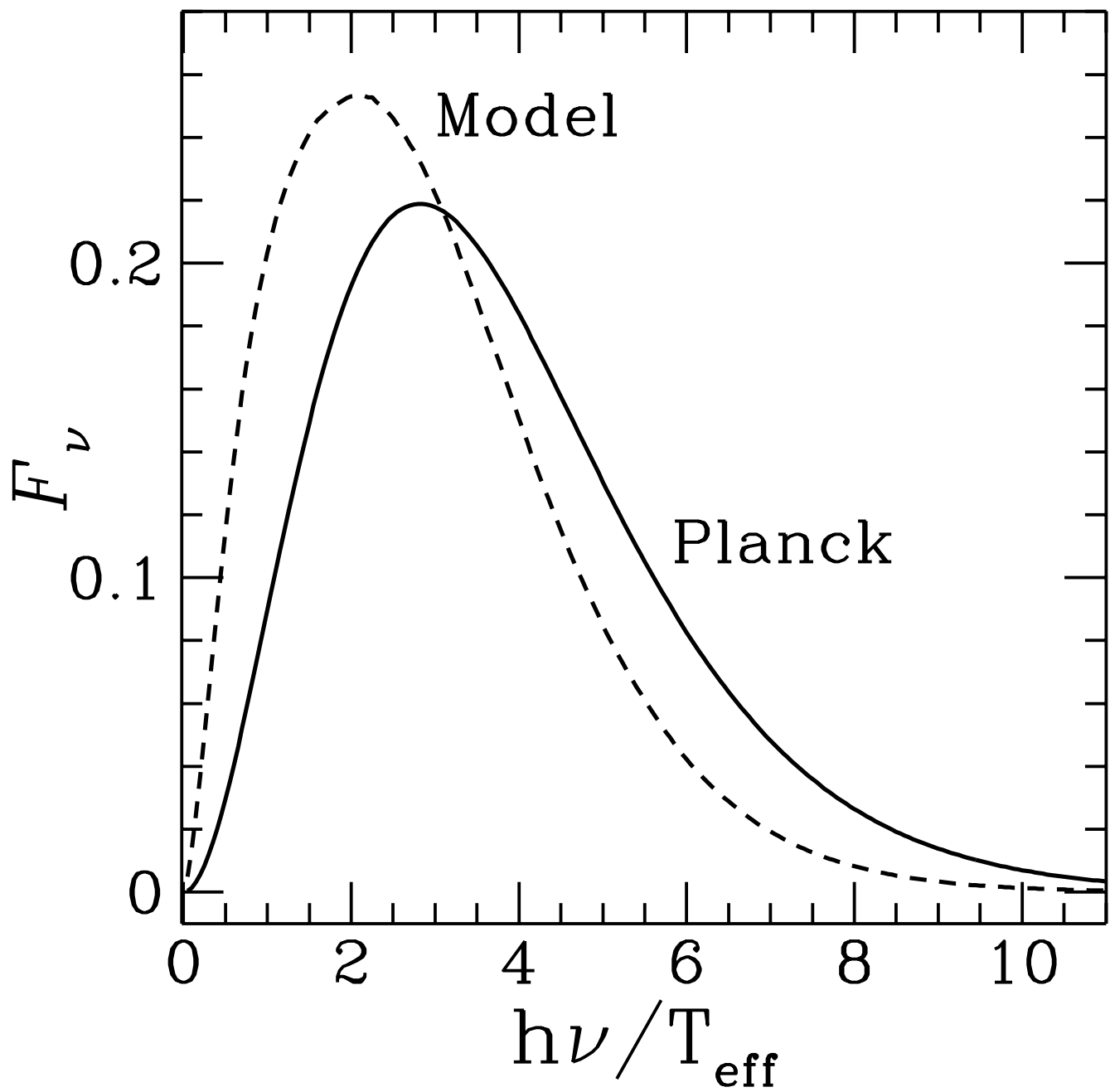
\bye